\documentclass[aps,prd,nofootinbib,psfig,graphicx,array]{revtex4}
\usepackage[colorlinks=true, pdfstartview=FitV, linkcolor=blue, citecolor=red, urlcolor=magenta]{hyperref}
\usepackage{graphicx}
\usepackage{latexsym}
\usepackage{amsmath}
\usepackage{amsfonts}
\usepackage{amssymb}
\usepackage{verbatim}

\renewcommand{\narrowtext}{\begin{multicols}{2} \global\columnwidth20.5pc}

\newcommand{\be}{\begin{equation}}
\newcommand{\ee}{\end{equation}}
\newcommand{\bea}{\begin{eqnarray}}
\newcommand{\eea}{\end{eqnarray}}

\newcommand{\pa}{\partial}

\newcommand{\remark}[1]{}

\newcommand{\ben}{\begin{eqnarray}}
\newcommand{\een}{\end{eqnarray}}

\begin{document}

\title{The radiatively corrected Kaluza-Klein masses in aether compactification}


\author{R. J. S. Oliveira}
\email{rafael.afram3@gmail.com}
\affiliation{Instituto de F\'{\i}sica, Universidade de Bras\'{i}lia,
Caixa Postal 04455, 70919-970, Bras\'{i}lia, DF, Brazil}

\author{M. A. Anacleto}
\email{anacleto@df.ufcg.edu.br}
\affiliation{Departamento de F\'{\i}sica, Universidade Federal de Campina Grande,\\
Caixa Postal 10071, 58429-900, Campina Grande, Para\'{\i}ba, Brazil}

\author{F. A. Brito}
\email{fabrito@df.ufcg.edu.br}
\affiliation{Departamento de F\'{\i}sica, Universidade Federal de Campina Grande,\\
Caixa Postal 10071, 58429-900, Campina Grande, Para\'{\i}ba, Brazil}
\affiliation{Departamento de F\'{\i}sica, Universidade Federal da Para\' iba,\\
Caixa Postal 5008, 58051-970, Jo\~ ao Pessoa, Para\'{\i}ba, Brazil}

\author{O. Holanda}
\email{netoholanda91@gmail.com}
\affiliation{ Centro de Matem\'atica, Computa\c c\~ao e Cogni\c c\~ao, Universidade Federal do ABC,  Avenida dos Estados 5001, 09210-580, Santa Terezinha, Santo Andr\' e - SP - Brazil. }

\author{E. Passos}
\email{passos@df.ufcg.edu.br}
\affiliation{Departamento de F\'{\i}sica, Universidade Federal de Campina Grande,\\
Caixa Postal 10071, 58429-900, Campina Grande, Para\'{\i}ba, Brazil}

\author{A. Pinzul}
\email{apinzul@unb.br}
\affiliation{Instituto de F\'{\i}sica, Universidade de Bras\'{i}lia,
Caixa Postal 04455, 70919-970, Bras\'{i}lia, DF, Brazil}



\begin{abstract}
We address the issue of  radiative corrections to Kaluza-Klein (KK) masses in five-dimensional QED supplemented by aether Lorentz-violating terms. Specifically, we compute the corrections to the KK photon masses from one fermion loop. In general, the KK masses receive radiative corrections due to breaking the five-dimensional Lorentz invariance by compactification. As we show, the presence of the additional Lorentz violating factor - an aether background, leads to the non-trivial modification of these corrections. This model may be of interest in addressing important phenomenological issues such as the relation between radiative corrected KK mass splitting  of a particular mode and uncertainties in the measurements and/or possible spatial variation of the fine-structure constant. For the recent data on the fine-structure constant, we find a KK mass splitting of magnitude $\sim 0.01$ MeV for the first excited Kaluza-Klein gauge boson at TeV scale. On the other hand, the large KK modes limit displays a very interesting phenomenon, showing the very special role of the aether in protecting the higher modes from the quantum corrections.

\end{abstract}
 \maketitle


\section{Introduction}

The aether compactification has been put forward sometime ago as an alternative to compactification in the presence of large extra dimensions without introduction of branes to control the influence of five-dimensional bulk space-time on the field content in our four-dimensional Universe. Differently from braneworlds, in this type of compactification there are no corrections to the four-dimensional Newtonian law \cite{Carroll:2008pk}. However the possibility of suppressing the Kaluza-Klein (KK) modes makes the mechanism efficient to guarantee a four-dimensional effective theory describing the physics of our Universe after the compactification. This is because Lorentz-violating aether fields along extra dimensions affect the conventional Kaluza-Klein compactification scheme, since their interactions play a fundamental role in the mass splitting of the KK towers: the mass spacings between different states are now modified, as we shall see later in greater details. In the case of large extra dimensions, even very high KK modes could, in principle, be accessible to a four-dimensional observer.

The stabilization of the extra dimension in aether compactification has been also studied in \cite{Chatrabhuti:2009ew}, where the effects of the aether field on the moduli stabilization mechanism were discussed. In this case the authors considered a space-like aether field aligned along the compact fifth direction with a Maxwell-type term. They show that an interplay between the Casimir energy due to massless and massive bulk fields can produce a potential for stabilization of the compact extra dimension.

Another important problem regarding KK masses concerns their radiative corrections. In \cite{Cheng:2002iz}, this issue was addressed by considering the one-loop radiative corrections to KK masses in five and six-dimensional theories. Such a computation has been shown to produce a finite result since the non-trivial winding number (corresponding to the compactification) contributions to Feynman loop diagrams are well-defined and cut-off independent even if higher dimensional theories are not renormalizable. Because this exactly corresponds to the part of loop diagrams that leads to the violation of the 5-dimensional Lorentz invariance induced by the presence of a compact dimension (as probed by IR or large-distance physics), it contributes to the correction of the KK masses. (See \cite{Rizzo:2005um} for a comprehensive study on Lorentz violation in extra dimensions \cite{PerezLorenzana:2005iv}.) On the other hand, for the zero winding number, i.e. for the short-distance physics, the contribution to loop diagrams is Lorentz invariant (and divergent). To isolate the finite Lorentz violating corrections from the divergent Lorentz invariant contributions one proceeds as follows: from every loop of the compactified theory one subtracts the corresponding loop of the uncompactified theory. Since the theories are equivalent in UV (short-distance) regime, the divergences are canceled. However, the KK mass corrections remain unchanged and finite due to the Lorentz invariance of the subtraction prescription.

Several studies combining the aether field, Lorentz and parity symmetry violation and radiative corrections in four and higher dimensional extended QED have been conducted in the literature. The aether-like Lorentz-violating term radiatively induced in extended Lorentz-violating four and five-dimensional QED has been previously addressed in \cite{Scarpelli:2013eya,Mariz:2016gqj}. The aether field modifications of the Stefan-Boltzmann law and the Casimir effect at both zero and finite temperature have been investigated in \cite{Santos:2017yov}. In \cite{CcapaTtira:2010ez}, the induction of a parity breaking term was found by computing the vacuum polarization tensor from a five-dimensional QED compactified in a magnetic flux background. Further models, such as cosmology in the presence
of dynamical timelike four-vector aether field in a four-dimensional theory (a sigma-model aether) and in the presence of aether constant field along the extra dimension in the inflationary Universe, have also been examined in \cite{Carroll:2009en}.

The main goal of the present study is to investigate the one-loop radiative corrections to the KK masses in the Lorentz-violating five-dimensional QED (QED$_5$) extended by a vector (aether) field with expectation value aligned along the extra dimension. To this end, we apply the subtraction prescription introduced in \cite{Cheng:2002iz} and briefly summarized above. More specifically, we compute the corrections to the KK photon masses coming from one fermionic loop. It is known, \cite{Cheng:2002iz}, that at this order the masses receive corrections even without the presence of aether. Our aim is to study the effect of the interaction with aether background. We will see that the one-loop radiative corrections get non-trivially modified, which could, in principle, lead to new phenomenological bounds on the Lorentz-violating parameters and measurements of the fine-structure constant. By considering the recent results of Ref.~\cite{Wilczynska:2020rxx}, i.e., $\Delta\alpha/\alpha=-2.18 \pm7.27 \times 10^{-5} $, and assuming the compactification radius $R\sim$ 1 mm, for Kaluza-Klein gauge bosons at TeV scale we found KK mass splitting uncertainty with the magnitude $\sim 0.01$ MeV for the first excited mode.

This paper is organized as follows: in section \ref{sec011}, we consider the spontaneous compactification of Maxwell-aether theory initially defined in 5D. We show that while the zero modes correspond to the standard 4D Maxwell and real scalar field with the modified kinetic part of the latter, the nonzero modes correspond, after the appropriate gauge fixing, to a family of non-interacting complex Proca fields. In section \ref{sec01}, we provide some basic information about the aether-QED extension in  5D. We describe the gauge symmetry of the model and define the one-loop polarization tensor. In section \ref{sec02}, using the modified fermionic propagator we compute radiative corrections to masses of Kaluza-Klein modes in the compactified QED$_5$ extended by the aether coupled to massless fermions. The exact integral form for the KK masses is obtained, which allows us to analyze different interesting limits, controlled by some effective coupling constant $\Lambda$: i) critical, $\Lambda =0$, that includes the case without aether studied in \cite{Cheng:2002iz}); ii) close to critical and iii) large KK, which is a very interesting limit, showing the very special role of the aether in protecting the higher modes from the quantum corrections. Also, the existence of a unitary bound is established. The section \ref{sec03} contains our conclusions.

\section{Compactification of Maxwell-Aether Theory}\label{sec011}

Let us consider the following five-dimensional effective action for  electromagnetic field extended by the lowest-order coupling to the aether background field $u^{a}$, given as \cite{Carroll:2008pk} (we use the metric with the mostly minuses signature)
\bea\label{em01}
S_{MA} = \int d^{5}x \Big[ -\frac{1}{4}  F_{ab}F^{ab} + \frac{1}{2 \mu_{A}^{2}} u^{a} u^{b} \eta^{cd}F_{ac} F_{bd}\Big] \ ,
\eea
where $x^a = (x^\mu , x^5)$ and $\mu = \overline{0,3}$.
The field strength tensor $F_{ab}$ is defined in terms of the potential $A_{a}$ by the usual relation:
\bea\label{em02}
F_{ab} = \pa_{a}A_{b} - \pa_{b} A_{a}\ .
\eea
The effective action, Eq.(\ref{em01}), is invariant under the gauge transformation: $\delta A^{a} = \pa^{a} \omega(x)$. We consider an aether background of the form:
\bea\label{em03-0}
u^{a} = (0, 0, 0, 0, v)\ ,
\eea
where the extra dimension coordinatized by $x^{5}$ is compactified on a circle of radius $R$. We correspondingly split the vector potential as $A^{a} =(A^{\mu}, A^{5}=\phi)$. This leads to
\bea\label{em03}
S_{MA} &=& \dfrac{1}{2\pi R}\int\!\!\!d^4x\!\!\!\int_{0}^{2\pi R}\!\!\!dx^5\bigg[-\dfrac{1}{4}F_{\mu\nu}F^{\mu\nu} - \dfrac{1}{2}(1 + \alpha_A^2)F_{5\mu}F^{5\mu}\bigg]  \nonumber\\
&=& \dfrac{1}{2\pi R}\int\!\!\!d^4x\!\!\!\int_{0}^{2\pi R}\!\!\!dx^5\bigg[-\dfrac{1}{4}F_{\mu\nu}F^{\mu\nu} + \dfrac{1}{2}(1 + \alpha_A^2)\big(\partial_{\mu}\phi\partial^{\mu}\phi - \partial^5A^{\mu}\partial_5A_{\mu}\big) - (1 + \alpha_A^2)\partial^{\mu}\phi\partial^5A_{\mu}\bigg]\ ,
\eea
where $\alpha_A = v/\mu_A$. Notice that the first term of the action (\ref{em03}) cannot be identified with the Maxwell action at four-dimensional spacetime, because the vector potential $A^{\mu}$ depends on the both coordinates, $x^{\mu}$ and $x^{5}$. By following the standard procedure adopted in studies of compactified models, we consider the Fourier expansions of the fields $A^\mu$ and $\phi$ as follows \cite{PerezLorenzana:2005iv}
\begin{subequations}
\begin{equation}\label{ema}
A^{\mu}(x^{\mu},x^5) = \sum_{n=-\infty}^{\infty}A_{(n)}^{\mu}(x)e^{i n x^5/R}\ ,
\end{equation}
\begin{equation}\label{emb}
\phi(x^{\mu},x^5) = \sum_{n=-\infty}^{\infty}\phi_{(n)}(x)e^{i n x^5/R}\ ,
\end{equation}
\end{subequations}
where the requirement of the single-valuedness of $A^a (x)$ leads to the quantization of $k_5$, $k_{5} = n/R$, $n\in \mathbb{Z}$. Now inserting the above expansions into Eq.(\ref{em03}), we obtain
\bea\label{em04}
S_{MA}  &=& \int\!\!d^4x\!\!\!\sum_{n=-\infty}^{\infty}\bigg[-\dfrac{1}{4}F_{(n)}^{\mu\nu}F_{(-n)\mu\nu} + (1 + \alpha_A^2)\bigg(\dfrac{1}{2}\partial_{\mu}\phi_{(n)}\partial^{\mu}\phi_{(-n)} + \dfrac{n^2}{2R^2}A_{(n)}^{\mu}A_{(-n)\mu} -\nonumber\\&& \dfrac{in}{R}A_{(-n)}^{\mu}\partial_{\mu}\phi_{(n)}\bigg)\bigg] =\nonumber\\&=&
\int\!\!\!d^4x\bigg\{-\dfrac{1}{4}F_{(0)}^{\mu\nu}F_{(0)\mu\nu} + \dfrac{1}{2}(1 + \alpha_A^2)\partial_{\mu}\phi_{(0)}\partial^{\mu}\phi_{(0)}  + \sum_{n=1}^{\infty}\bigg[ -\dfrac{1}{2}F_{(n)}^{\mu\nu}F_{(-n)\mu\nu} + (1 + \alpha_A^2)\bigg( \partial^{\mu}\phi_{(n)}\partial_{\mu}\phi_{(-n)} +\nonumber\\&&   \dfrac{n^2}{R^2}A_{(n)}^{\mu}A_{(-n)\mu}  +\dfrac{in}{R}A_{(n)}^{\mu}\partial_{\mu}\phi_{(-n)}  - \dfrac{in}{R}A_{(-n)}^{\mu}\partial_{\mu}\phi_{(n)}\bigg)\bigg]\bigg\}\ .
\eea
Since $A_{a}$ is a real quantity, we have that $A^{\mu}_{(-n)}= A^{*\mu}_{(n)}$ and $\phi_{(-n)} = \phi^{*}_{(n)}$. Thus, the Eq.(\ref{em04}) can be rewritten more conveniently:
\bea\label{em05}
S_{MA}  &=& \int\!\!\!d^4x\bigg\{-\dfrac{1}{4}F_{(0)}^{\mu\nu}F_{(0)\mu\nu} + \dfrac{1}{2}(1 + \alpha_A^2)\partial_{\mu}\phi_{(0)}\partial^{\mu}\phi_{(0)} + \sum_{n=1}^{\infty}\bigg[ -\dfrac{1}{2}F_{(n)}^{\mu\nu}F_{(n)\mu\nu}^* + \nonumber\\&&(1 + \alpha_A^2)\dfrac{n^2}{R^2}\bigg(A_{(n)}^{\mu} - \dfrac{iR}{n}\partial_{\mu}\phi_{(n)}\bigg)\bigg(A_{(n)\mu}^* + \dfrac{iR}{n}\partial_{\mu}\phi_{(n)}^*\bigg)\bigg]\bigg\} \ ,
\eea
where now all the fields are defined on the 4-dimensional space-time. Notice that the aether background does not affect the zero mode term at four dimensions, corresponding to the Maxwell theory, but re-scales the kinetic part of the scalar sector. For $n\ne 0$, the effective action (\ref{em05}) takes the form of the infinite sum of decoupled Stueckelberg actions for the complex fields $A_{(n)}^{\mu}$ and $\phi_{(n)}$. As it is well known, this theory is $U(1)$ gauge invariant if $A_{(n)}^{\mu}$ transforms as the usual connection and $\phi_{(n)}$ is in the affine representation, i.e. under $U(1)$ they transform as
\bea\label{Affine-gauge}
A_{(n)}^{\mu}\rightarrow A_{(n)}^{\mu} + \partial^\mu \omega_{(n)}\ ,\ \ \phi_{(n)} \rightarrow \phi_{(n)} - \frac{in}{R} \omega_{(n)} \ .
\eea
As usual, we can fix the gauge, $\phi_{(n)} = 0$, and arrive at the action for the family of the non-interacting massive Proca fields with the KK masses given by $m_{KK(n)}^2\equiv (1 + \alpha_A^{2}) \frac{n^2}{R^2}$,
\bea\label{em06-0}
S^{KK}_{MA} = \int\!d^4x \sum_{n=1}^{\infty}\bigg[ -\dfrac{1}{2}F_{(n)}^{\mu\nu}F_{(n)\mu\nu}^* + (1 + \alpha_A^2)\dfrac{n^2}{R^2}A_{(n)}^{\mu} A_{(n)\mu}^* \bigg] \ ,
\eea
so the KK masses are enhanced by a factor $(1 + \alpha_A^2)$ compared to the usual compactification without the aether background. This leads to the following modified dispersion relations \cite{Rizzo:2005um}:
\bea\label{em06}
E^{2} = k^2 + (1 + \alpha_A^{2}) \frac{n^2}{R^2},\,\,\,\,\,\,\,\,\,\,\,k= |\vec{k}|.
\eea
Note that though we have fixed the gauge of the KK part of the theory, the full resulting action, which is a sum of $n=0$ Maxwell/scalar part of (\ref{em05}) and $n\ne 0$ part (\ref{em06-0}), still has the residual $U(1)$ symmetry under which the $n\ne 0$ KK modes transform trivially (see also the discussion after Eq.(\ref{eqa4-0})).

As a trivial classical analysis of the action (\ref{em06-0}), let us look at the propagation of electromagnetic waves with the modified dispersion relation (\ref{em06}), by investigating the group and phase velocities. As usual, we shall assume that the photon follows the group velocity defined as
\bea\label{CR18}
v_{g}= \frac{\pa E}{\pa k} = \frac{1}{\sqrt{1 + (1 + \alpha_{A}^{2})   \big(n/ k R \big)^{2}}}\ .
\eea
On the other hand, the phase velocity is defined as $v_{p} = E/k$, i.e.,
\bea\label{CR19}
v_{p} = \sqrt{1 + (1 + \alpha_{A}^{2})   \big(n/ k R \big)^{2}}\ .
\eea
From here we easily find the relationship between the Eqs.(\ref{CR18}) and (\ref{CR19})
\bea\label{CR20}
\frac{v_{p} - v_{g}}{v_{g}} =  (1 + \alpha_{A}^{2})   \big(n/ k R \big)^{2}\ ,
\eea
in complete agreement with the classical Rayleigh's formula,  $\frac{v_{p}-v_{g}}{v_{g}} =  - \frac{k}{v_{g}} \frac{d v_{p}}{dk}$. Notice that for any value of $\alpha_{A}$, this implies $v_p  >  v_g $, which means a normal dispersion medium.

\section{One-loop quantum correction of aether Lorentz-violating QED$_{5}$} \label{sec01}

The main purpose of this section is to introduce the model within which we will compute the radiative corrections to KK photon masses (given by the dispersion relation (\ref{em06})). This model is given by the aether Lorentz-violating massless QED in five dimensions described by the following effective action:
\bea\label{eqa4}
S^{(5)}_{QEDA} = \int d^{5}x \Big( i\bar\psi\gamma^{a}\pa_{a}\psi  - \frac{i}{\mu_{\psi}^{2}}u_{a}u^{b}\bar\psi\gamma^{a}\pa_{b}\psi- e_{5}\bar\psi \gamma^{a}A_{a}\psi \Big)\ ,
\eea
where the 5D $\gamma$-matrices and the metric are defined as
\bea\label{gamma}
\gamma^a = (\gamma^\mu , i\gamma^5),\;  \{\gamma^\mu , \gamma^\nu\}=2\eta^{\mu\nu},\;\gamma^5 = i\gamma^0\gamma^1\gamma^2\gamma^3\;{\rm and} \;\eta^{\mu\nu}=\mathrm{diag}(1,-1,-1,-1) .
\eea
For the background (\ref{em03-0}) the action (\ref{eqa4}) takes the form
\bea\label{eqa4-0}
S^{(5)}_{QEDA} = \int d^{5}x \Big( i(\bar\psi\gamma^{\mu}\pa_{\mu}\psi + i\bar\psi (1+\alpha_\psi^2)\gamma^{5}\pa_{5}\psi ) - e_{5}\bar\psi \gamma^{a}A_{a}\psi \Big)\ ,
\eea
where $\alpha_\psi^2 := \frac{v^2}{\mu_{\psi}^{2}}$. The full action is the sum of (\ref{eqa4-0}) and the action from the previous section (\ref{em05}) (in the Stueckelberg gauge, cf. (\ref{em06-0})). The action (\ref{eqa4-0}) is invariant under the gauge transformation
\bea\label{eqa5}
&&\delta A_{a} = \frac{1}{e_{5}} \pa_{a}\omega(x)\ ,\nonumber\\&&
\delta\psi(x)= -i\omega(x)\psi(x)\ ,\nonumber\\&&
\delta\bar\psi(x)= i\omega(x)\bar\psi(x)\ ,
\eea
where $\pa_{5} \omega(x)=0$. This symmetry is compatible with the residual $U(1)$ gauge symmetry of (\ref{em05}) left after fixing the Stueckelberg gauge, i.e. while the higher KK modes, $n\ne 0$, of $A_\mu$ and the only surviving, $n=0$, mode of $\phi\equiv A_5$ are the scalars under this symmetry, the zero mode of $A_\mu$ is the usual $U(1)$ gauge boson and $\psi$ belongs to the fundamental representation.

Now we trivially show that the effect of the interaction with the aether background in (\ref{eqa4-0}) could be taken into account exactly by bringing the model to the form, which will look as in the ``aether-less'' scenario. This is achieved by defining the set of the deformed $\gamma$-matrices:
\bea\label{gamma-deformed}
\tilde{\gamma}^a = (\gamma^\mu , i(1+\alpha_\psi^2)\gamma^5)\ \mathrm{with}\ \{\tilde{\gamma}^a , \tilde{\gamma}^b\}=2g^{ab}\ ,\ \mathrm{and}\ g^{ab}=\mathrm{diag}(1,-1,-1,-1,-(1+\alpha_\psi^2))\ .
\eea
Then (\ref{eqa4-0}) takes the usual form
\bea\label{eqa4-1}
S^{(5)}_{QEDA} = \int d^{5}x \Big( i\bar\psi\tilde{\gamma}^{a}\pa_{a}\psi - e_{5}\bar\psi \gamma^{a}A_{a}\psi \Big)\ .
\eea
So the whole effect of the aether-fermion interaction is reduced to the difference between the physical and effective (amplified by the factor of $(1+\alpha_\psi^2)$) scales of the fifth dimension\footnote{This  effect is in the spirit of the spectral geometry where the physical geometry is determined by some relevant Dirac operator. In this sense, the effect of the coupling to aether is similar to matter coupling to the Horava-Lifshitz gravity in the low energy regime, where the similar re-scaling effectively happens for the whole space part of the metric and not just for one direction \cite{Pinzul:2014mva}.} and it is taken into account \textit{exactly} by the minimal modification of the fermionic propagator:
\bea\label{propagator}
\tilde{S}(k):= \frac{i}{\tilde{\gamma}^a k_a}\ .
\eea
Of course, this scale mismatch will have non-trivial consequences after the compactification.

Note that in (\ref{eqa4-1}) we have two types of the $\gamma$-matrices: one is entering the propagator (\ref{propagator}) and the other is responsible for the vertex of the perturbation series. Actually, it is easy to see that the minimal action (\ref{eqa4}) is not the most general one. One should also include the term $e_u u_a u^b \bar\psi \gamma^{a}A_{b}\psi \equiv - i e_u v^2 \bar\psi \gamma^{5}A_{5}\psi$, where $e_u$ is another arbitrary parameter. But this term can be handled in exactly the same way as above by defining the second set of the deformed matrices, $\tilde{\tilde{\gamma}}^a = (\gamma^\mu , i(1+e_u / e_5)\gamma^5)$. So the result will again look as in (\ref{eqa4-1}) but with a modified vertex. This modification will have an effect on the scalar, $n=0$, mode of $\phi\equiv A_5$ (recall that all the higher modes of $A_5$ were ``eaten'' by the Stueckelberg gauge).

\begin{figure}[h]
 \raisebox{-0.5cm}{\includegraphics[angle=0,scale=0.6]{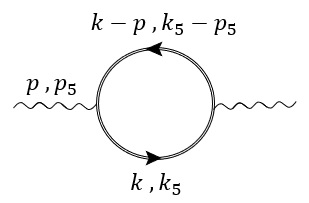}}
 \caption{The one-loop contribution to the polarization tensor.}\label{loop-diagram}
\end{figure}

Within the model, just defined, we would like to study the corrections to the KK photon masses induced by the interaction with the fermionic sector. These corrections will come from the part of the vacuum polarization tensor, $i\Pi^{ab}$, proportional to the metric. We will be interested in the corrections to the vector KK modes, i.e. given by $i\Pi^{\mu\nu}$. The contribution to the mass of the scalar KK zero mode will come from $i\Pi^{55}$ and could be calculated in complete analogy. (So for us it is not important which $\gamma$-matrices enter the vertex, because $\tilde{\tilde{\gamma}}^\mu \equiv \gamma^\mu$.) The first non-trivial contribution to $i\Pi^{\mu\nu}$ comes from the diagram on Fig.\ref{loop-diagram}. For the convenience of the future calculations, from now on we adopt the following notations: $k,\ p$, etc with indices or without will be used to denote four-momenta, while for the momenta in the fifth direction the index 5 will be used explicitly.

To calculate the contribution of the diagram on Fig.\ref{loop-diagram}, we use the standard Feynman rules with the only exception that we should use the modified massless fermionic propagator (\ref{propagator})
\begin{subequations}
\bea\label{RF1}
\raisebox{-0.3cm}{\includegraphics[angle=0,scale=0.3]{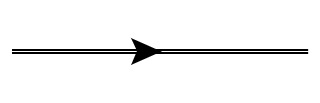}}
&& \tilde{S}(k)=\frac{i}{k_{a}\tilde{\gamma}^a }\ ;
\eea
\bea\label{RF3}
 \raisebox{-0.3cm}{\includegraphics[angle=0,scale=0.3]{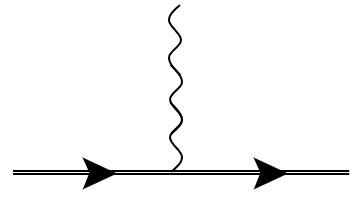}}
 && V(e_{5})= - i e_{5}\gamma^{a}\ .
\eea
\end{subequations}
Then, to this order, the 4D part of the photon vacuum polarization tensor is given by (as usual, ``-'' comes from the fermionic loop and there is still no compactification!)
\begin{equation}\label{polarisation1}
i\Pi^{\mu\nu} = -e_5^2\int\!\!\! \dfrac{d^5k}{(2\pi)^5}\text{tr}\bigg[(-i\gamma^{\mu})\dfrac{i}{k\!\!\!/ + i(1+\alpha_{\psi}^2)\gamma^5k_5}(-i\gamma^{\nu})\dfrac{i}{(k\!\!\!/ - p\!\!\!/) + i(1+\alpha_{\psi}^2)\gamma^5(k_5-p_5)}\bigg]\ .
\end{equation}

The compactification on a circle of the radius $R$ along the fifth direction is standardly taken into account by passing from the integral over $k_5$ to the sum over $k_5 = \frac{m}{R}$, $m\in \mathbb{Z}$
\bea\label{integral-sum}
\int \frac{d^{5}k}{(2\pi)^{5}} \to \frac{1}{2 \pi R} \sum_{k_{5}}  \int \frac{d^{4}k}{(2\pi)^{4}}\ ,
\eea
so that the expression for the polarization in the compactified case becomes (after defining the effective 4D electric charge $e^{2}=e^{2}_{5}/(2\pi R) $)
\begin{equation}\label{polarisation2}
i\Pi^{\mu\nu} = -e^2\sum_{k_5}\int\!\!\! \dfrac{d^4k}{(2\pi)^4}\dfrac{\text{tr}\big[\gamma^{\mu}\big(k\!\!\!/ + i(1+\alpha_{\psi}^2)\gamma^5k_5\big)\gamma^{\nu}\big(k\!\!\!/ - p\!\!\!/ + i(1+\alpha_{\psi}^2)\gamma^5(k_5-p_5)\big)\big]}{\big(k^2-(1+\alpha_{\psi}^2)^2 k_5^2\big)[(k-p)^2-(1+\alpha_{\psi}^2)^2(k_5-p_5)^2]}\ .
\end{equation}

Before calculating this one-loop contribution to the vacuum polarization (which we do in the next section), it is useful to compare (\ref{polarisation2}) to the case without Lorentz breaking aether background, discussed in \cite{Cheng:2002iz}.

First of all, of course, in both cases the corresponding diagram is divergent and this divergence is dealt with in the same way - by the subtraction of the uncompactified result. This is done via the Poisson summation formula, which essentially trades the winding and KK numbers. This is explained in more details in the next section.

Then we have two obvious differences with the ``aether-less'' case: 1) The modified dispersion relation (\ref{em06}) for the on-shell KK photons and 2) the modified fermionic propagator (\ref{propagator}) (which is, of course, the consequence of the modification of the dispersion relation analogous to (\ref{em06}) but now for the fermion sector). We will see that while the former leads to some technical complications in the calculations, the latter (combined with the former) has a rather interesting physical consequence, relating the seemingly independent coupling constants $\alpha_A$ and $\alpha_\psi$.

\section{Contribution Induced by Aether compactification}\label{sec02}
In this section, we provide the results of our calculations for one-loop corrections via the subtraction prescription and apply the obtained results to find the corrections to the masses of the KK photons.

As we mentioned in the previous section, to render the expression (\ref{polarisation2}) finite, one has just to subtract from it the uncompactified result (\ref{polarisation1}). \cite{Cheng:2002iz} This procedure has a very clear physical meaning (and standardly used in quantum field theory on curved backgrounds): because the divergence is due to the bad UV behaviour of the diagram, it should be exactly the same as in the uncompactified case (UV physics does not ``know'' that the space is compact in one direction). In practice this is done by rewriting the KK sum in (\ref{polarisation2}) in terms of the winding modes using the Poisson summation formula \cite{Cheng:2002iz}
\bea\label{PRI}
\frac{1}{2\pi R} \sum_{m=-\infty}^{+\infty} F(m/R) \to \sum_{n=-\infty}^{+\infty}  f(2 \pi R n)\ ,
\eea
where the functions  $f(x)$ and $F(k)$ are related by Fourier transformation
\bea\label{FT}
f(x) = {\cal F}^{-1}\{F(k)\} = \int_{-\infty}^{+\infty} \frac{dk}{2\pi} e^{i k x} F(k)\ ,
\eea
and throwing away the term corresponding to the zero winding number. (Here, $m$ corresponds to the KK number, while $n$ - to the winding number.) One can think of this as discarding the contribution, which does not probe the compactness of the extra dimension. Really, the $n=0$ contribution to the sum is identical to the integral over the uncompactified extra dimension:
\bea\label{FT01}
f(0) = \int_{-\infty}^{+\infty} \frac{d k_{5}}{2\pi} F(k_{5})\ .
\eea
So, it is obvious that in this way, the subtraction prescription simply removes the divergent $n=0$ term and the remaining terms in the sum, i.e., $n\neq 0$, are exactly the ones that keep the information about the compact fifth dimension. As we will explicitly see below, they are all finite and the sum is convergent.

Before applying the described subtraction prescription to our case let us first do the standard manipulations with Eq.(\ref{polarisation2}): calculate the trace over spinorial indices and use the Feynman parametrization.
The result for the trace is given by
\bea\label{trace}
N^{\mu\nu}&:=&\text{tr}\big[\gamma^{\mu}\big(k\!\!\!/ + i(1+\alpha_{\psi}^2)\gamma^5k_5\big)\gamma^{\nu}\big(k\!\!\!/ - p\!\!\!/ + i(1+\alpha_{\psi}^2)\gamma^5(k_5-p_5)\big)\big] \nonumber\\
&:=&4\Big[k^{\mu}\big(k^{\nu} - p^{\nu}\big) + k^{\nu}\big(k^{\mu} - p^{\mu}\big) - \eta^{\mu\nu}k\cdot\big(k-p\big) + \eta^{\mu\nu}(1+\alpha_{\psi}^2)^2k_5\big(k_5-p_5\big)\Big]\ .
\eea

To deal with the denominator of (\ref{polarisation2}) we, as usual, use the Feynman trick
\bea\label{Feynman}
\frac{1}{AB} = \int\limits_0^1 \frac{dx}{(x A + (1-x) B)^2}\ \ \mathrm{for\ all}\ A,B \ .
\eea
Then (\ref{polarisation2}) becomes
\begin{equation}\label{polarisation3}
i\Pi^{\mu\nu} = -e^2\int_{0}^{1}\!\!\! dx\sum_{k_5}\int\!\!\! \dfrac{d^4k}{(2\pi)^4}\dfrac{\tilde{N}^{\mu\nu}}{\big[k^2 - (1+\alpha_{\psi}^2)^2k_5^{\prime 2} + x(1-x)\big(p^2 - (1+\alpha_{\psi}^2)^2p_5^2\big)\big]^2}\ ,
\end{equation}
where we defined $k_5^{\prime 2} := k_5 - p_5x$ and $\tilde{N}^{\mu\nu}=\left.{N}^{\mu\nu}\right|_{k^\mu \rightarrow k^\mu + p^\mu x}$. In this formula, $p^2$ is a 4D momentum of an on-shell KK photon corresponding to $p_5 = l/R$, $l\in \mathbb{Z}$. To this order of approximation, we can use the zero order on-shell condition (\ref{em06}), i.e. $p^2 = (1+\alpha_A^2)p_5^2$. Plugging this into (\ref{polarisation3}), discarding the terms linear in $k^\mu$ and replacing $k^\mu k^\nu$ with $\frac{1}{4} k^2 \eta^{\mu\nu}$ (due to the obvious symmetries of the integral) and, finally, re-scaling all the momenta by $R$ (so now, $p_5 = l$ and $k'_5 = m-lx$, $l,m\in \mathbb{Z}$) we get
\bea\label{polarisation4}
i\Pi^{\mu\nu} = &-&\frac{4e^2}{R^2}\sum_{m\in\mathbb{Z}}\int_{0}^{1}\!\!\! dx\int\!\!\! \dfrac{d^4k}{(2\pi)^4}\Big\{\eta^{\mu\nu}\Big[-\dfrac{k^2}{2} + x(1-x)\big(1+\alpha_A^2 - (1+\alpha_{\psi}^2)^2\big)p_5^2 + \notag \\
&+& (2x-1)(1+\alpha_{\psi}^2)^2p_5k_5^{\prime} + (1+\alpha_{\psi}^2)^2k_5^{\prime 2}\Big] - 2x(1-x)p^{\mu}p^{\nu}\Big\}\times \notag \\ &\times &\dfrac{1}{\big[k^2 - (1+\alpha_{\psi}^2)^2 k_5^{\prime 2} + x(1-x)\big(1 + \alpha_A^2 - (1+\alpha_{\psi}^2)^2\big)p_5^2\big]^2}\ .
\eea

Before we proceed, let us take a look at the obtained expression. First of all, as a trivial check, it is easily verified that for $\alpha_A = \alpha_\psi = 0$ we have the correct expression for the ``aether-less'' case \cite{Cheng:2002iz}. But for the general value of the coupling constants there is a very new feature. While the factor of $(1+\alpha_{\psi}^2)^2$ in front of $k_5^2$ in the denominator is just a consequence of the aforementioned modified fermionic dispersion relation, the factor of $\big(1 + \alpha_A^2 - (1+\alpha_{\psi}^2)^2\big)$ in front of $p_5^2$ is much more interesting. Because $x(1-x)$ is non-negative for all $x\in [0,1]$, the sign of $x(1-x)\big(1 + \alpha_A^2 - (1+\alpha_{\psi}^2)^2\big)p_5^2$ depends only on the sign of $\big(1 + \alpha_A^2 - (1+\alpha_{\psi}^2)^2\big)$. If this is positive, one can always find a high enough KK momentum $p_5$, so this term would dominate $(1+\alpha_{\psi}^2)^2(k_5 - p_5 x)^2$ for any fixed $k_5^{2}$ and some $x$. This will lead to an imaginary pole in (\ref{polarisation4}) signalling about problems with unitarity. So, we should conclude that the theory has a \textit{unitary bound} on the coupling constants:\footnote{Of course, this could be seen directly from (\ref{polarisation2}): if $(1+\alpha_{\psi}^2)^2 < 1 + \alpha_A^2$ then $\tilde{S}(k-p)$ has an imaginary pole for any fixed $k_5$ and high enough $p_5$. This also shows that the unitary bound does not depend on the subtraction procedure.}
\bea\label{unitary-bound}
(1+\alpha_{\psi}^2)^2 \geq 1 + \alpha_A^2 \ .
\eea
From now on, to ensure one-loop unitarity, we will assume that (\ref{unitary-bound}) is satisfied. (Of course, in the ``aether-less'' case, it is satisfied trivially.)

It is well-known that the corrections to the KK photon masses will come from the part of (\ref{polarisation4}) proportional to the metric, $\eta^{\mu\nu}\Pi_{\eta}$, which after the usual Wick rotation, $k_0 \rightarrow ik_0$, is given by
\bea\label{polarisation-metric}
\Pi_{\eta} = -\frac{4e^2\Lambda_\psi^2}{R^2}\sum_{m}\int_{0}^{1}\!\!\! dx\int\!\!\! \dfrac{d^4\bar{k}_E}{(2\pi)^4}\times \bigg\{\dfrac{\dfrac{\bar{k}_E^2}{2} - x(1-x)\Lambda^2 p_5^2 + (2x-1)p_5k_5^{\prime} + k_5^{\prime 2}}{\big[\bar{k}_E^2 + k_5^{\prime 2} + x(1-x)\Lambda^2 p_5^2\big]^2}\bigg\}\ ,
\eea
where in our notation:
$\bar{k}_{E}= k_{E}/ \Lambda_{\psi}$ with  $\Lambda_\psi := 1+\alpha_{\psi}^2$ and we introduced the ``effective'' coupling constant,
\bea\label{cc01}
\Lambda^2 := -\frac{1+\alpha_A^2 - \Lambda_\psi^2}{\Lambda_\psi^2}\,.
\eea
It is clear that the whole effect of $\Lambda_\psi$ is reduced to the multiplicative modification of the electric charge, while $\Lambda$ really controls how the quantum corrections deviate from the ``aether-less'' case. Note that the unitary bound (\ref{unitary-bound}) is important to ensure that $\Lambda^2$ is non-negative. It is easy to see that $\Lambda^2 \in [0,1)$.

With the help of the formula
\bea\label{CR09}
\frac{1}{A^{r}} = \frac{1}{( r - 1) !} \int_{0}^{\infty} d\ell \,\ell^{r - 1}\, e^{-A \ell}
\eea
the equation (\ref{polarisation-metric}) can be transformed into the form permitting integration over $\bar{k}_E$, the result being (after the variable change $\ell=1/t$)\footnote{Note that the unitary bound (\ref{unitary-bound}) guarantees that the transformation (\ref{CR09}) makes sense: if (\ref{unitary-bound}) is violated the application of (\ref{CR09}) in (\ref{polarisation-metric}) would be divergent.}
\bea\label{polarisation-metric1}
\Pi_\eta = -\dfrac{e^2 \Lambda_\psi^2}{4\pi^2R^2}\sum_{m}\int_{0}^{1}\!\!\! dx\int_{0}^{\infty}\!\!\! dt\bigg\{1 + \dfrac{(2x-1)p_5k_5^{\prime}}{t} - \dfrac{x(1-x)\Lambda^2 p_5^2}{t}
+ \dfrac{k_5^{\prime 2}}{t}\bigg\}e^{-[k_5^{\prime 2} + x(1-x)\Lambda^2 p_5^2\big]/t}\ .
\eea

At this point, we are finally ready to apply the Poisson summation formula (\ref{PRI}). For this we have to calculate inverse Fourier transformation of the integrand in (\ref{polarisation-metric1}) with respect to $k_5$. The relevant Fourier integrals are standard and are given by
\bea\label{CR11}
&& \mathcal{F}^{-1}\big\{e^{ -[k_5^{2} + x(1-x)\Lambda^2 p_5^2\big]/t}\big\} = \sqrt{\dfrac{t}{4\pi}}e^{-x(1-x)\Lambda^2 p_5^2/t}~e^{-\tfrac{y^2t}{4}}\ ,\nonumber\\
&& \mathcal{F}^{-1}\big\{k_5 e^{-[ k_5^{2} + x(1-x)\Lambda^2 p_5^2\big]/t}\big\} = \Big(\dfrac{iyt}{2}\Big)\sqrt{\dfrac{t}{4\pi}}e^{-x(1-x)\Lambda^2 p_5^2/t}~e^{-\tfrac{y^2t}{4}}\ ,\nonumber\\
&& \mathcal{F}^{-1}\big\{k_5^2 e^{-[k_5^{2} + x(1-x)\Lambda^2 p_5^2\big]/t}\big\} = \Big(\dfrac{t}{2}-\dfrac{t^2y^2}{4}\Big)\sqrt{\dfrac{t}{4\pi}}e^{-x(1-x)\Lambda^2p_5^2/t}~e^{-\tfrac{y^2t}{4}}\ ,\nonumber\\
&& \mathcal{F}^{-1} \{ F\{ k^{\prime\,2}_{5} = k_{5} - x p_{5} \} \} = f(y) e^{i x y p_{5}}\ .
\eea
Using these results in the Poisson summation formula (\ref{PRI}), we can transform the sum in (\ref{polarisation-metric1}), which goes over the KK modes, $m$, into the sum over the winding numbers, $n$
\begin{align}\label{polarisation-metric2}
\Pi_\eta = -\dfrac{\Lambda_\psi^2e^2}{2\pi R^2}\sum_{y=-\infty}^{\infty}\int_{0}^{1}\!\!\! dxe^{ixyp_5}\int_{0}^{\infty}\!\!\! dt\sqrt{\dfrac{t}{4\pi}}e^{-x(1-x)\Lambda^2 p_5^2/t}~e^{-\tfrac{y^2t}{4}}\bigg\{\dfrac{3}{2} + i\Big(x-\dfrac{1}{2}\Big)yp_5 - \dfrac{y^2t}{4} - \dfrac{x(1-x)\Lambda^2 p_5^2}{t}\bigg\}\ ,
\end{align}
where $y=2\pi n$. This can be simplified to the form maximally resembling the case $\Lambda = 0$:
\begin{align}\label{polarisation-metric3}
\Pi_\eta = -\dfrac{\Lambda_\psi^2e^2}{2\pi R^2}\sum_{y=-\infty}^{\infty}\int_{0}^{1}\!\!\! dxe^{ixyp_5}\int_{0}^{\infty}\!\!\! dt\sqrt{\dfrac{t}{4\pi}}e^{-x(1-x)\Lambda^2 p_5^2/t}~e^{-\tfrac{y^2t}{4}} \bigg\{i\Big(x-\dfrac{1}{2}\Big)yp_5 - \dfrac{2x(1-x)\Lambda^2 p_5^2}{t}\bigg\}\ .
\end{align}
In the transition of the Eq.(\ref{polarisation-metric2}) to Eq.(\ref{polarisation-metric3}) we used the the observation that under the integral, i.e. up to a surface term, $ 3/2 - y^{2} t/ 4 = - x(1-x) \Lambda^{2} p_{5}^{2}/t $.

Finally, the integral over $t$ is calculated with the help of the following standard result:
\bea\label{nice-integral}
\int\limits_0^\infty e^{-\left( a x^2 + \frac{b}{x^2} \right)} dx = \frac{1}{2}\sqrt{\frac{\pi}{a}} e^{-2\sqrt{ab}}\ ,\ \ \mathrm{for\ any}\ a,b >0\ .
\eea
The result of this integration is given by
\begin{align}\label{polarisation-final}
\Pi_\eta = -\dfrac{\Lambda_\psi^2 e^2}{2\pi R^2}\sum_{y=-\infty}^{\infty}\int_{0}^{1}\!\!\! dxe^{ixyp_5}\bigg\{i(2x-1)\dfrac{yp_5}{|y|^3}\bigg[1+{\sqrt{x(1-x)}\Lambda|p_5||y|}\bigg] - \dfrac{2x(1-x)\Lambda^2|p_5|^2}{|y|}\bigg\}e^{-\sqrt{x(1-x)}\Lambda|p_5||y|}\ .
\end{align}
Though this could be re-written in several equivalent ways, it does not seem that any of them have some advantage compared to the form (\ref{polarisation-final}). (See, though, the discussion of the large KK limit below.) At the moment, we were unable to calculate (\ref{polarisation-final}) analytically for an arbitrary value of $\Lambda$. So, we will analyze and test this result from several different points of view and in different limits.

First of all, let us verify that the mass corrections do not depend on the sign of $p_5$, i.e. that (\ref{polarisation-final}) is an even function of $p_5$. This is quite easy to see by sending $p_5 \rightarrow - p_5$ and using the symmetry of the sum over the winding modes with respect to $y \rightarrow -y$. So as a first trivial check, we have
\bea
\delta{m}^{2}_{KK} \equiv \Pi_\eta (p_5) = \Pi_\eta (-p_5)\ .
\eea

As was discussed after Eq.(\ref{eqa5}), the zero mode KK photon, corresponds to the massless, $p_5=0$, $U(1)$ gauge boson. So, we should expect that this residual gauge symmetry is respected by the quantization and this mode does not receive any quantum correction to the mass. This is indeed the case, because it is trivial from (\ref{polarisation-final}) that
 \begin{align}
\Pi_\eta (p_5 = 0) = 0\ .
\end{align}

To study the massive KK modes, we have to, as was discussed above, perform a minimal subtraction by discarding the $y=0$ mode from the sum in (\ref{polarisation-final}) (note that the $y=0$ term is indeed the only divergent one). As it was explained, this corresponds to subtraction of the polarization tensor for the uncompactified model. The next, trivial, observation (and some kind of a consistency check) is that for the case of $\Lambda = 0$ the integration in (\ref{polarisation-final}) is trivial and we reproduce exactly the trivially re-scaled result of \cite{Cheng:2002iz} without aether (recall that $y=2\pi n$)
\bea\label{CR13}
\delta{m}^{2}_{KK} \equiv \Pi_\eta (\Lambda =0) &=&
- \frac{ e^{2}\Lambda_{\psi}^{2}}{2\pi R^{2}} \sum_{n = 1}^{+\infty}
\frac{2}{|2\pi n |^{3}} \nonumber\\&=& - \frac{\Lambda_{\psi}^{2} e^{2} \zeta(3)}{4\pi^4 R^{2}} \ ,
\eea
where $\zeta (x)$ is the standard Riemann zeta-function. Actually, this result is not entirely trivial. The saturated unitary bound $\Lambda = 0$ imposes a relation between the coupling constants: $1+\alpha_A^2 = (1+\alpha_\psi^2)^2$. Usually this would signal about some symmetry in an underlying fundamental theory. It is worth reminding that our result is  \textit{exact} in these coupling constants (the perturbation parameter is the electric charge, $e$ or, rather, the fine structure constant). So the exact form of the quantum corrections (\ref{CR13}) could be a good way to test such a symmetry.

This observation shows the importance of the study of the quantum corrections close to the saturation of the unitary bound, i.e. for $\Lambda \ll 1$. It is difficult to study this limit directly from Eq.(\ref{polarisation-final}), because $\Lambda$ effectively enters exponent in the combination $\Lambda |y|$, which can be arbitrarily large for high terms in the sum. It is more convenient to work directly with (\ref{polarisation-metric3}), where $\Lambda$ is a nice expansion parameter for any $y$ (actually, the effective expansion parameter is $\Lambda |p_5|$, so the obtained result will be valid only for the lower KK modes, such that $\Lambda |p_5|\ll 1$, see also below)
\bea\label{polarisation-expanded}
\Pi_\eta = -\dfrac{\Lambda_\psi^2e^2}{2\pi R^2}\sum_{n=-\infty}^{\infty}\int_{0}^{1}\!\!\! dxe^{ixyp_5}\int_{0}^{\infty}\!\!\! dt\sqrt{\dfrac{t}{4\pi}}e^{-\tfrac{y^2t}{4}}\bigg\{i\Big(x-\dfrac{1}{2}\Big)yp_5 - \dfrac{x(1-x)\Lambda^2 p_5^2}{t}\left( i\Big(x-\dfrac{1}{2}\Big)yp_5 +2\right)\bigg\} + \mathcal{O}(\Lambda^4)\ ,
\eea
which could be easily integrated over $t$ and then over $x$. Then the result up to $\mathcal{O}(\Lambda^4)$ takes the form
\bea\label{polarisation-expanded1}
\delta{m}^{2}_{KK} \equiv \Pi_\eta &=& -\dfrac{\Lambda_\psi^2e^2}{2\pi R^2}\sum_{n\ne 0}\int_{0}^{1}\!\!\! dxe^{ixyp_5}\bigg\{i\Big(2x-1\Big)\frac{yp_5}{|y|^3} - \dfrac{x(1-x)\Lambda^2 p_5^2}{|y|}\left( i\Big(x-\dfrac{1}{2}\Big)yp_5 +2\right)\bigg\}=\notag \\
&=&-\frac{\Lambda_\psi^2e^2}{2\pi R^2}\sum_{n\ne 0}\frac{2+ 8\Lambda^2}{(2\pi n)^3}\equiv - \frac{\Lambda_{\psi}^{2} e^{2} \zeta(3)}{4\pi^4 R^{2}}\left( 1+4\Lambda^2 \right)\ .
\eea
So we see that the whole effect of $\Lambda\ne 0$ is again just an overall factor and the result does not depend on the number of a KK mode as in the case $\Lambda = 0$, Eq.(\ref{CR13}). But as we mentioned above, this is true only for the lowest modes, such that $\Lambda |p_5|\ll 1$. We will see that in the opposite limit the effect is much more drastic.

To study the limit of the large KK number, it is convenient to use the exact final result (\ref{polarisation-final}). For the large KK number (by this we mean that $\Lambda |p_5| \gg 1$) the exponent in (\ref{polarisation-final}) will be dominated by the value of $x$ close to the end-points of the interval $[0,1]$. Using the symmetry of the sum with respect to $y$ and with the help of the change of the variable, $s:=\sqrt{x(1-x)} \Lambda |p_5||y|$, which is well-defined for $x\in [0,1/2]$, one can easily bring (\ref{polarisation-final}) to the following equivalent form (still no approximation!)
\begin{align}\label{polarisation-final-1}
\Pi_\eta = -\dfrac{\Lambda_\psi^2 e^2}{\pi R^2}\sum_{n\ne 0}\int\limits_{0}^{\frac{1}{2}\Lambda|p_5||y|}\!\!\! e^{iyp_5\frac{1-\sqrt{1-4s^2/\Lambda^2p_5^2 y^2}}{2}}\frac{2s ds}{|y|^5\Lambda^2 p_5^2}\bigg\{ -i(1+s)yp_5 - \frac{2s^2}{\sqrt{1-4s^2/\Lambda^2p_5^2 y^2}} \bigg\}e^{-s}\ .
\end{align}
In $\Lambda |p_5| \gg 1$ regime, we can extend the integration limit to infinity, this will introduce an error of order of $\mathcal{O}\left(e^{-\frac{1}{2}\Lambda|p_5||y|}\right)$. Then we can expand the integrand in $1/\Lambda |p_5|$ up to the first non-trivial term
\begin{align}\label{polarisation-Large-KK}
\Pi_\eta = -\dfrac{\Lambda_\psi^2 e^2}{\pi R^2}\sum_{n\ne 0}\int\limits_{0}^{\infty}\!\!\! \frac{2ds}{|2\pi n|^5\Lambda^2 p_5^2}\bigg\{ \frac{1}{\Lambda^2}s^3(1+s) - 2s^3 \bigg\}e^{-s} + \mathcal{O}\left(\frac{1}{\Lambda^4 p_5^4}\right)\ .
\end{align}
Now the integral and the sum can be trivially calculated leading to the large KK number corrections
\bea\label{polarisation-Large-KK-1}
\delta{m}^{2}_{KK} \equiv \Pi_\eta &=& -\frac{\Lambda_\psi^2e^2}{2\pi R^2}\frac{1}{\Lambda^2 p_5^2}\left( \frac{2}{\Lambda^2}(\Gamma(3)+\Gamma(4)) - 4\Gamma(3) \right)\sum_{n\ne 0}\frac{1}{(2\pi n)^5}\nonumber\\&\equiv& -\frac{\Lambda_\psi^2e^2 \zeta(5)}{4\pi^6 R^2}\frac{1}{\Lambda^2 p_5^2}\left( \frac{2}{\Lambda^2} - 1 \right).
\eea
Note that this result is nonperturbative in $\Lambda$, in particular, it does not make sense to take $\Lambda \rightarrow 0$ (for $\Lambda = 0$ the change of variable leading to (\ref{polarisation-final-1}) is degenerate). The result (\ref{polarisation-Large-KK-1}) shows very important role of the aether in the case of the non-critical coupling, $\Lambda \ne 0$. While in the case of the saturated unitary bound (including the ``aether-less'' case), $\Lambda = 0$, the mass corrections described by Eq.(\ref{CR13}) are universal, i.e. independent of the KK momentum $p_5$, away from criticality, the same is true only for the lowest KK modes, Eq.(\ref{polarisation-expanded1}). For the higher KK excitations, the quatum corrections to the masses are suppressed by the inverse powers of $p_5^2$. This could, in principle, be of a great experimental importance, if an experiment is designed to probe the higher KK modes.

{
At this point, we develop some analyses of possible phenomenological consequences of the obtained results. In the case of the small effective coupling constant, $\Lambda\ll 1$, Eq.(\ref{polarisation-expanded1}), leads to the following modification of the dispersion relation (\ref{em06}):
\bea\label{ph01}
k_{\mu} k^{\mu} = {\bar m}_{KK}^{2} =  \Lambda_{\psi}^{2}\big(1 - \Lambda^{2}  \big) \frac{n^2}{R^2} - \lambda_{(n)} \Lambda_{\psi}^{2}\big(1 + 4 \Lambda^{2}  \big) \frac{n^2}{R^2}\ .
\eea
Note that the corrections to the Kaluza-Klein masses are modified (compared to \cite{Cheng:2002iz}) by a factor $ \lambda_{(n)} \Lambda_{\psi}^{2}\big(1 + 4 \Lambda^{2}  \big)$ (with $\lambda_{(n)} = e^{2} n^{-2}\zeta(3)/4\pi^4$), which involves the Lorentz-violating aether parameter and the fine-structure constant, which may be important for the precision phenomenology. In writing (\ref{ph01}), we trivially re-wrote Eq.(\ref{em06})
\bea\label{ph02}
k_{\mu} k^{\mu} = {m}_{KK}^{2} =  \Lambda_{\psi}^{2}\big( 1 - \Lambda^{2}  \big) \frac{n^2}{R^2}
\eea
using the definition (\ref{cc01}). For a related discussion of experimental constraints on aether parameter, see Refs. \cite{Carroll:2009en, Jacobson:2004ts}. Below, we shall focus on the bounds on varying fine-structure constant.

To this end, let us compare the KK masses given in Eq.(\ref{ph01}) and Eq.(\ref{ph02})
\bea\label{rad-bare-KK}
\left| \frac{\bar{m}^{2}_{KK}-m^{2}_{KK}}{m^{2}_{KK}}\right|=\lambda_{(n)}+{\cal O}(\Lambda^2),
\eea
where $\lambda_{(n)}$ can be re-written in terms of the fine-structure constant $\alpha$, i.e.,  $\lambda_{(n)} = \alpha\, \zeta(3)/n^2 \pi^{3}$. Note that as it was discussed above, up to the order of ${\cal O}(\Lambda^2)$, this result is the same as in the case of the compactification without aether. Now we can relate the radiative corrections to the KK masses with the small uncertainty $\Delta\alpha$ of the bare fine-structure constant \cite{Martins:2017yxk}. Substituting $\alpha\to\alpha+\Delta\alpha$ into (\ref{rad-bare-KK}) we find, up to leading term (or in the critical limit, $\Lambda = 0$), and for $n=1$
\bea\label{rad-bare-KK-2}
\left| \frac{\bar{m}_{KK}^2-m_{KK}^2}{m_{KK}^2}\right| &\simeq& \left(\frac{\alpha\zeta(3)}{\pi^3}\right)+\left(\frac{\alpha\zeta(3)}{\pi^3}\right)\frac{\Delta\alpha}{\alpha}\nonumber\\
&\simeq& 0.00028+0.00028\, \frac{\Delta\alpha}{\alpha}.
\eea
In the last step of the equation above we have used $\Big({\alpha\,\zeta(3)}/{\pi^3}\Big)\approx 0.00028$, $\zeta(3)\approx 1.20205$, for the bare value $\alpha\approx 1/137$.

It is important to notice that the effect of the fine structure uncertainty in (\ref{rad-bare-KK-2}) can be \textit{completely} masked by the $\Lambda^2$ terms discarded in (\ref{rad-bare-KK}). Really, up to these terms, we have
\bea\label{rad-bare-KK2}
\left| \frac{\bar{m}^{2}_{KK}-m^{2}_{KK}}{m^{2}_{KK}}\right|=\lambda_{(n)}(1 + 3\Lambda^2)+{\cal O}(\Lambda^4)\ .
\eea
This shows that if $\Lambda$ is not identically zero, but instead $\Lambda^2 \sim \frac{\Delta\alpha}{\alpha}$, it would be impossible to tell whether the effect is due to $\Delta\alpha$ or some non-critical $\Lambda$ or both. To disentangle these two contributions, one would need to go either to the higher KK analysis (see below) or to use some other experiment to put independent bounds.

For the case when $\Lambda|p_5|\gg 1$ (\ref{polarisation-Large-KK-1}) (because $\Lambda\in [0,1)$, this necessarily corresponds to higher KK modes), we have, using the same definition of $\lambda_{(n)}$ as above,
\bea\label{ph01-2}
k_{\mu} k^{\mu} = {\bar m}_{KK}^{2} =  \Lambda_{\psi}^{2}\big( 1 - \Lambda^{2}  \big) \frac{n^2}{R^2} - \frac{\lambda_{(n)}}{\pi^2}\frac{\zeta(5)}{\zeta(3)}\frac{\Lambda_{\psi}^{2}}{\Lambda^4p_5^2}\left( 2 -\Lambda^{2}\right)\frac{n^2}{R^2}
\eea
and using (\ref{ph02})
we find
\bea\label{rad-bare-KK1}
\left| \frac{\bar{m}_{KK}^2-m_{KK}^2}{m_{KK}^2}\right|=\lambda_{(n)}\frac{\zeta(5)}{\zeta(3)} \frac{(2-\Lambda^2)}{(1-\Lambda^2)}\left(\frac{1}{\pi \Lambda}\right)^2\left(\frac{1}{\Lambda |p_5|}\right)^2.
\eea
Note that in (\ref{rad-bare-KK1}) there are two competing possibilities. On the one hand, because $\Lambda |p_5|\gg 1$ we have that the relative radiative corrections are suppressed by the factor $\frac{1}{(\Lambda |p_5|)^2}$ (as the absolute corrections (\ref{polarisation-Large-KK-1})). On the other hand, if we are in the strong non-critical regime, i.e. $\Lambda$ is close to $1$ (which corresponds to $\Lambda^2_{\psi}\gg 1+\alpha^2_A$) then the term $(1-\Lambda^2)$ in the denominator of (\ref{rad-bare-KK1}) might become important rending the relative radiative corrections to be large. All of this shows the importance of the large KK limit as it reveals the non-trivial role of the aether (somewhat hidden for the lower modes). In principle, this should allow us to distinguish between the critical models (the ``aether-less'' model being a special case), for which the results (\ref{rad-bare-KK}), (\ref{rad-bare-KK-2}) are valid for any value of $p_5$, and the non-critical ones. At the same time this would allow to disentangle the effects due to the variations of the fine structure and $\Lambda\ne 0$.

Let us do some estimations for the critical case (\ref{rad-bare-KK-2}). For the radius of the fifth dimension $R\sim$ 1 mm, $v\sim M_{P}$ and $\mu_A\sim 1$ TeV, we find $m_{KK}\sim 1$ TeV for the mode $n=1$ \cite{Carroll:2008pk}. The spatial variation and uncertainty of the fine-structure constant, i.e., $\Delta\alpha/\alpha=-2.18 \pm7.27 \times 10^{-5} $ has been tested very recently even in large redshifts as in the early Universe \cite{Wilczynska:2020rxx}. In the present scenario one might impose bounds on this uncertainty if we assume that higher changes of $\alpha$ appear in the direction of the fifth dimension and we know the uncertainty of radiatively corrected masses KK masses. Alternatively, if we assume the well-know deviations of $\alpha$ we can estimate the uncertainty of such KK masses. In the example above, the bare KK mass is $m_{KK}\sim 1$ TeV, then the uncertainty in the mass splitting has the magnitude $|\Delta\bar{m}_{KK}|\sim 0.01$ MeV for the first excited mode. In a Randall-Sundrum model \cite{Randall:1999ee}, the Kaluza-Klein excitations of  gauge bosons $V_{KK}$ \cite{Djouadi:2007eg} are expected to be excited in the Large Hadron Collider (LHC) via the Drell-Yan process, $pp\to q\bar{q}\to V_{KK}\to Q\bar{Q}$, where $q$ and $Q$ are initial and final heavy fermions, such as top and bottom quarks. The KK gauge boson decay into light fermions such as leptons and lower generation of quarks cannot be neglected in general \cite{Angelescu:2017jyj}, such that light KK mass splitting can be probed. Thus, inside the aforementioned radiative corrected KK mass splitting, there exists the possibility to accomplish an excess of KK gauge boson decays into well known light fermions such as  muon neutrino, $m_{\nu_\mu}<0.2$ MeV, electron neutrino, $m_{\nu_e} < 2$ eV, and so on, as a consequence of spatial variation and uncertainty of the fine-structure constant.

\section{Conclusions}\label{sec03}

In the present study we have considered the aether compactification in a five-dimensional quantum electrodynamics. The effective four-dimensional theory presents excitations with radiatively corrected Kaluza-Klein masses coming from the 5D gauge bosons. These KK masses depend fundamentally on both aether parameter (and the corresponding coupling constants) and the fine-structure constant $\alpha$. The aether compactification modifies the mass spacing between different Kaluza-Klein states as a consequence of the interaction with the aether field aligned along the compact fifth dimension. We investigate the radiative corrections to the Kaluza-Klein masses by computing the vacuum polarization in the presence of a compact fifth dimension with the compactification radius $R$. A finite result is found by considering a prescription where from every fermion loop of the compactified theory the corresponding fermion loop of the uncompactified theory is subtracted. We would like to stress that  the calculations are non-perturbative in the coupling constants, the perturbation is done only with respect to the fine structure constant. This allows us to see the unitary bound, unseen in the perturbative in $\alpha_\psi$ approach. The exact integral form for the radiative corrections to the masses is obtained, which allowed to analyze different interesting limits: i) critical, $\Lambda =0$ (this includes the case without aether studied in \cite{Cheng:2002iz}); ii) close to critical; iii) large KK - this limit is very interesting, showing the very special role of the aether in protecting the higher modes from the quantum corrections, among other effects.

Our results demonstrate a potential possibility to address several important phenomenological issues. One interesting route is to relate the deviations of the measurements of the fine-structure constant to KK mass splitting. The variation and uncertainty of the fine-structure constant has been tested very recently even in large redshifts as in the early Universe \cite{Wilczynska:2020rxx}. In the present scenario, in the critical case of $\Lambda = 0$, we can, in principle, impose bounds on this uncertainty if we assume that higher changes of $\alpha$ appear in the direction of the fifth dimension. By considering the results of Ref.~\cite{Wilczynska:2020rxx}, i.e., $\Delta\alpha/\alpha=-2.18 \pm7.27 \times 10^{-5} $ and assuming $R\sim$ 1 mm, for Kaluza-Klein gauge bosons at TeV scale we found KK mass splitting uncertainty with the magnitude $\sim 0.01$ MeV for the first excited mode. For the non-critical case, to disentangle the effects due to $\Delta\alpha$ and $\Lambda \ne 0$, one has to go to the higher KK modes (or use other independent experiments). Further investigations concerning the possibility of the decays of such KK gauge bosons into light fermions, associated with KK mass splitting related to spatial variation and uncertainties of the fine-structure constant, seems to be phenomenologically interesting from both particle physics and astrophysics point of view.

\eject

\acknowledgments

We would like to thank CNPq, CAPES and CNPq/PRONEX/FAPESQ-PB (Grant no. 165/2018),  for partial financial support. MAA, FAB and EP acknowledge support from CNPq (Grant nos. 306962/2018-7, 312104/2018-9, 304852/2017-1). RJSO was supported by CAPES PhD fellowship.


\end{document}